\documentclass[reprint, amsmath,amssymb,aps,prb, superscriptaddress]{revtex4-2}
\usepackage{graphicx}
\usepackage{dcolumn}
\usepackage{bm}
\DeclareMathOperator{\Tr}{Tr}
\usepackage{xcolor}
\usepackage{longtable}
\usepackage{colortbl}%
  
\usepackage{booktabs,tabularx}
\usepackage{dcolumn}
\usepackage{hyperref}

\hypersetup{
    colorlinks = true,
    citecolor = {blue},
    linkcolor = {blue},
    urlcolor  = {blue},
}
\let\newcolumn=\newpage
\renewcommand{\newpage}{\if@firstcolumn\newcolumn\null\fi \newcolumn}

\usepackage{array}
\newcolumntype{P}[1]{>{\centering\arraybackslash}p{#1}}
\newcolumntype{M}[1]{>{\centering\arraybackslash}m{#1}}

\usepackage{tikz}
\newcommand*\circled[1]{\tikz[baseline=(char.base)]{
            \node[shape=circle,draw,inner sep=2pt] (char) {#1};}}

\begin{document}

\title{Acoustic signatures of the field-induced electronic-topological transitions in YbNi$_4$P$_2$}

\author{E.-O. Eljaouhari}
\affiliation{Institut für Mathematische Physik, Technische Universität Braunschweig,
Mendelssohnstraße 3, 38106 Braunschweig, Germany}
\affiliation{Université de Bordeaux, CNRS, LOMA, UMR 5798, 33400 Talence, France}
\author{B. V. Schwarze}
\affiliation{Hochfeld-Magnetlabor Dresden (HLD-EMFL) and Würzburg-Dresden Cluster of Excellence ctd.qmat,
 Helmholtz-Zentrum Dresden-Rossendorf, 01328 Dresden, Germany}
\author{K. Kliemt}
\affiliation{Physikalisches Institut, Johann Wolfgang Goethe-Universität, D-60438 Frankfurt am Main, Germany}
\author{C. Krellner}
\affiliation{Physikalisches Institut, Johann Wolfgang Goethe-Universität, D-60438 Frankfurt am Main, Germany}
\author{F. Husstedt}
\affiliation{Hochfeld-Magnetlabor Dresden (HLD-EMFL) and Würzburg-Dresden Cluster of Excellence ctd.qmat,
 Helmholtz-Zentrum Dresden-Rossendorf, 01328 Dresden, Germany}
 \affiliation{Institut für Festkörper- und Materialphysik, TU Dresden, 01062 Dresden, Germany}
\author{J. Wosnitza}
\affiliation{Hochfeld-Magnetlabor Dresden (HLD-EMFL) and Würzburg-Dresden Cluster of Excellence ctd.qmat,
 Helmholtz-Zentrum Dresden-Rossendorf, 01328 Dresden, Germany}
\affiliation{Institut für Festkörper- und Materialphysik, TU Dresden, 01062 Dresden, Germany}
\author{S. Zherlitsyn}
\affiliation{Hochfeld-Magnetlabor Dresden (HLD-EMFL) and Würzburg-Dresden Cluster of Excellence ctd.qmat,
Helmholtz-Zentrum Dresden-Rossendorf, 01328 Dresden, Germany}
\author{G. Zwicknagl}
\affiliation{Institut für Mathematische Physik, Technische Universität Braunschweig,
Mendelssohnstraße 3, 38106 Braunschweig, Germany}
\affiliation{Max Planck Institute for Chemical Physics of Solids, 01187 Dresden, Germany}
\author{J. Sourd}
\affiliation{Hochfeld-Magnetlabor Dresden (HLD-EMFL) and Würzburg-Dresden Cluster of Excellence ctd.qmat,
Helmholtz-Zentrum Dresden-Rossendorf, 01328 Dresden, Germany}

\date{\today}\begin{abstract}
We investigated the magnetoelastic properties of an YbNi$_4$P$_2$ single crystal at low temperatures under magnetic fields directed along the crystallographic [001] axis. We report a series of strong anomalies in the sound velocity, which is consistent with the cascade of electronic-topological transitions reported previously for this compound. In particular, we identify the vanishing of a small orbit on the Fermi surface, associated with a quantum-oscillation frequency of 34 T. Furthermore, the different transitions are better resolved with acoustic modes of particular symmetry. Using a microscopic model adapted to the strongly correlated electronic structure of YbNi$_4$P$_2$, we describe our results by inspecting realistic electron-phonon couplings in reciprocal space for each acoustic mode. This shows how the $k$ selectivity of ultrasound experiments allows to investigate Fermi-surface reconstructions in strongly correlated electronic systems
\end{abstract}

\maketitle

\section{\label{sec:level1}Introduction}

The Fermi surface is a central concept to understand the physical properties of correlated-electron systems \cite{dugdale2016life}. Its shape and topology often dictate the structure of the exotic phases that can emerge from instabilities of the Fermi-liquid regime at low temperature. This is, for example, the case for the spin-density waves in CeCu$_2$Si$_2$ \cite{stockert2004nature}, and for the unconventional superconductivity in iron-based materials \cite{hirschfeld2016using}.

While resulting from the precise chemistry of a given compound through its crystal structure, chemical bonding, and number of conduction electrons, the Fermi-surface shape and topology can change drastically upon varying an external control parameter, leading to an electronic-topological transition (ETT) \cite{blanter1994theory}. This phenomenon is a natural consequence of the underlying crystal lattice: the electronic energy bands have to be periodic in reciprocal space, which enforces the existence of critical points in the band structure (maxima, minima or saddle points). The crossing of such critical point through the Fermi level then automatically leads to a change of connectivity of the Fermi surface, and, thus, an ETT. 

Historically, the concept of ETT was developed by Lifshitz in the 1960s in case of a transition induced by pressure \cite{lifshitz1960anomalies}. The first experimental study was reported a few years later in thallium \cite{brandt1966influence}. Yet, the most common route to realize an ETT is the use of chemical substitution as control parameter, adding extra electrons or holes to the system. This method was applied successfully in the 1980s for Li$_{1-x}$Mg$_x$ \cite{egorov1983thermopower} and is still very popular today, for instance, in the iron-based superconductors Ba$_{1-x}$K$_x$Fe$_2$As$_2$ \cite{xu2013possible} and Ba(Fe$_{1-x}$Co$_x$)$_2$As$_2$ \cite{liu2011importance}. 

A third way to induce an ETT is to apply an external magnetic field and take advantage of the Zeeman effect in order to fill or empty the electronic bands. In ordinary metals, the Fermi energy is of the order of electron-volt and, thus, an ETT is difficult to reach with magnetic fields available in laboratories. However, in heavy-fermion systems, the strong electronic correlations generate renormalized flat bands close to the Fermi level, leading to effective Fermi energies three orders of magnitude smaller. Thus, in these systems the Zeeman energy from a moderate magnetic field is sufficient to induce an ETT, as observed at 7.8 T in CeRu$_2$Si$_2$ \cite{daou2006continuous, boukahil2014lifshitz}, at 28 T in CeIrIn$_5$ \cite{aoki2016field}, at 3.4, 9.3, and 11 T in YbRh$_2$Si$_2$ \cite{naren2013lifshitz, pfau2013interplay, pourret2013magnetic}, and at a few fields below 18 T in YbNi$_4$P$_2$ \cite{pfau2017cascade}.

In order to detect an ETT, a rich portfolio of experimental techniques is available. Transport measurements such as the magnetoresistance and Hall effect are commonly used \cite{naren2013lifshitz, pfau2013interplay}, and, in particular, thermoelectric-power experiments are very sensitive to such transitions \cite{pourret2013magnetic, pfau2017cascade}. One can also probe an ETT by measuring the magnetic susceptibility \cite{tokiwa2004suppression} or magnetostriction \cite{pfau2013interplay, pfau2017cascade}, for example. Finally, the study of quantum oscillations can also reveal Fermi-surface reconstructions such as the appearance of a new pocket, as in FeSe$_{1-x}$S$_x$ \cite{coldea2019evolution} and in CeIrIn$_5$ \cite{aoki2016field}. 

However, these probes can only provide limited information about the Fermi-surface transformation associated with the ETT. Only the area of the orbit can be extracted from a quantum-oscillation measurement, for example, and neither the shape of the pocket nor its location in the Brillouin zone is directly accessible. Angle-resolved photoemission spectroscopy (ARPES) can then give very valuable insights into the location of the ETT in the reciprocal space. This technique has been recently used to characterize ETT in Ba(Fe$_{1-x}$Co$_x$)$_2$As$_2$ \cite{liu2011importance} and Ba$_{1-x}$Sr$_x$Ni$_2$As$_2$ \cite{narayan2023potential}. Raman spectroscopy also permits probing specific regions in the reciprocal space depending on the light polarization. It has been utilized to investigate the ETT in the cuprate superconductor Bi$_2$Sr$_2$CaCu$_2$O$_{8+\delta}$ \cite{benhabib2015collapse}. Yet, these two techniques are limited to rather high temperatures and zero or moderate magnetic fields.

In this paper, we studied the sequence of ETT in YbNi$_4$P$_2$ through electron-phonon scattering processes by performing ultrasound measurements. Such experiments have the advantage of being compatible with very low temperatures as well as high magnetic fields. Ultrasound has already been shown to be sensitive to ETT, such as the metamagnetic transition in CeRu$_2$Si$_2$ \cite{yanagisawa2002ultrasonic}. Furthermore, in some cases it permits exploring the reciprocal-space structure of correlated electron systems, such as the superconducting gap in cuprate and ruthenate superconductors \cite{walker2001electron}.

We analyze our ultrasound data by introducing a microscopic model adapted to the strongly correlated electronic structure of YbNi$_4$P$_2$. We first derive the evolution of Fermi surfaces induced by the magnetic field, which reproduce the multiple ETT detected. Secondly, we determine realistic electron-phonon couplings in reciprocal space depending on the acoustic-wave symmetry. Comparing these couplings to the observed anomalies in the sound velocity, we show how ultrasound experiments permit a meticulous exploration of the ETT in reciprocal space.

\section{Crystal and electronic structure}\label{sec2}

\begin{figure}[t!]
    \centering
    \includegraphics[width=\linewidth]{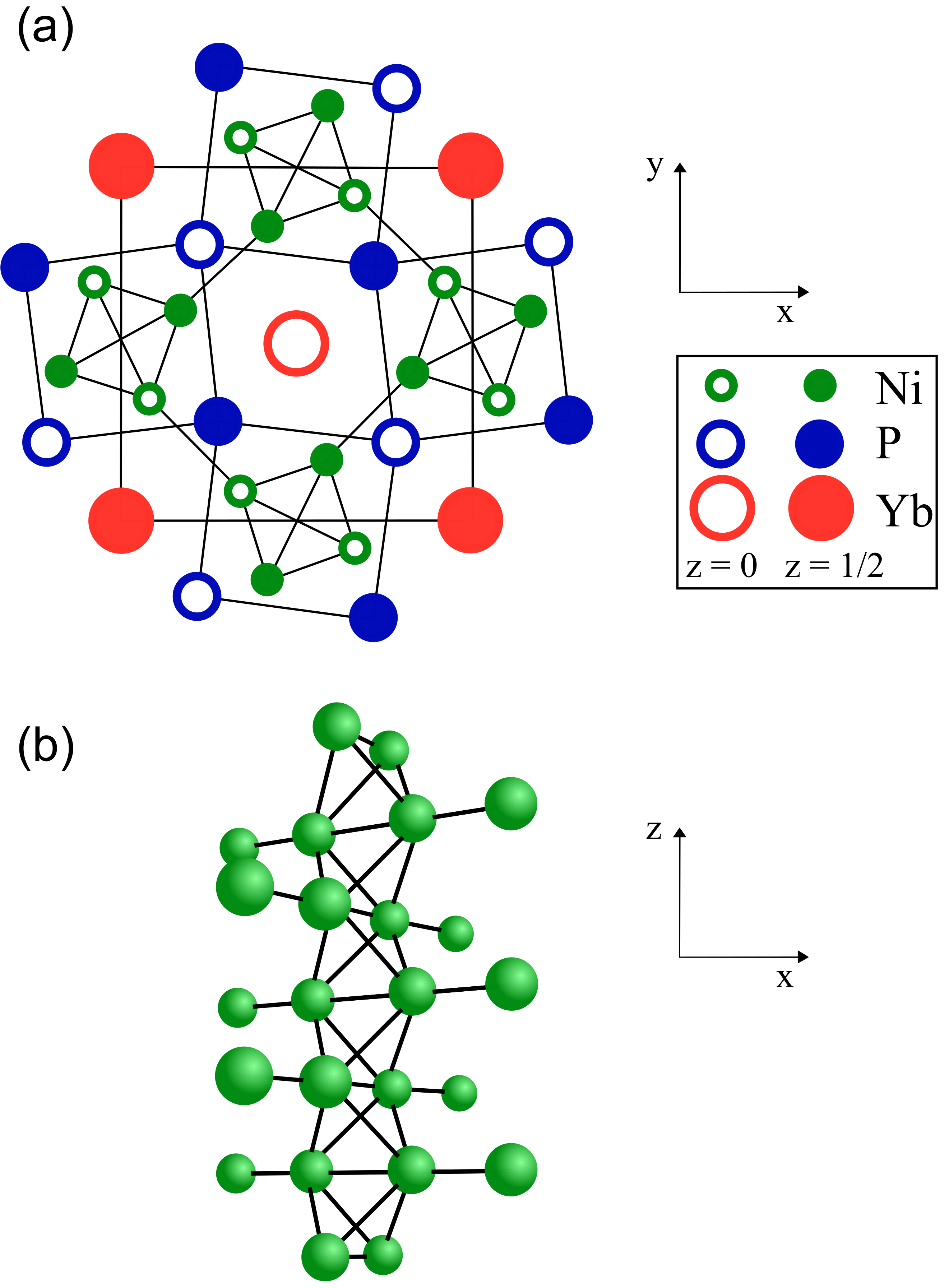}
    \caption{ (a) Projection of the YbNi$_4$P$_2$ crystal structure on the (001) plane. (b) Connections within the Ni$_4$ clusters leading to the quasi-1D behavior of YbNi$_4$P$_2$. This figure is adapted from \cite{le1993electronic}.}\label{fig0} 
\end{figure}

YbNi$_4$P$_2$ crystallizes in the tetragonal ZrFe$_4$Si$_2$-type structure (space group $P4_2/mnm$) shown in Fig. \hyperref[fig0]{\ref*{fig0}(a)}. According to Ref. \onlinecite{le1993electronic}, the main feature of this structure is the presence of tetrahedral Ni$_4$ clusters, which are fused together along the [001] direction and form linear chains [Fig. \hyperref[fig0]{\ref*{fig0}(b)}], leading to a strong quasi-one-dimensional (1D) anisotropy. It has been proposed in Refs. \cite{krellner2011ferromagnetic, steppke2013ferromagnetic} that the 1D behavior of YbNi$_4$P$_2$ might play a decisive role in explaining the presence of a ferromagnetic quantum critical point. This is a unique feature of this compound that cannot be explained by state-of-the-art theories for 2D or 3D metallic quantum ferromagnets \cite{brando2016metallic}. However, the observation of multiple field-induced ETT suggests that the electronic structure of YbNi$_4$P$_2$ is much richer than a collection of 1D sheets \cite{pfau2017cascade}.

Due to its rather complex unit cell and the presence of strongly correlated $4f$ holes, the detailed electronic structure of YbNi$_4$P$_2$ is hardly reproduced by ab-initio approaches. Up to now, most predictions are based on $f$-core DFT calculations of the non-magnetic analogue LuNi$_4$P$_2$ with the lattice parameters of YbNi$_4$P$_2$. These calculations show the presence of 1D sheets \cite{krellner2011ferromagnetic} as well as some sheets of mixed dimensionality between 1D, 2D, and 3D. In particular, the presence of small ``tube-like''  2D Fermi surfaces has been proposed to explain the effect of the magnetic-field direction on the ETT in Ref. \cite{karbassi2018anisotropic}. Recently, the presence of both 1D and mixed-dimensionality sheets has been confirmed by ARPES experiments \cite{dai2025electronic}.

In this work, we extend the existing electronic-structure calculations and incorporate the correlated $4f$ holes arising from the Kondo effect at low temperatures, as well as the magnetic field, as described in the following.

\section{Experimental Techniques}\label{sec2}

We grew single crystals of YbNi$_4$P$_2$ using the Czochralski method, as described in more detail in Ref. \cite{kliemt2016crystal}. We checked the phase purity by powder x-ray diffraction on crushed single crystals and using energy-dispersive x-ray spectroscopy. We selected a crystal of dimensions 1.0 mm $\times$ 1.0 mm $\times$ 2.6 mm, with the longest direction of 2.6 mm along the [001] axis. The orientation of the single crystal was determined using a Laue camera with x-ray radiation from a tungsten anode.

\begin{figure*}
    \centering
    \includegraphics[width=\linewidth]{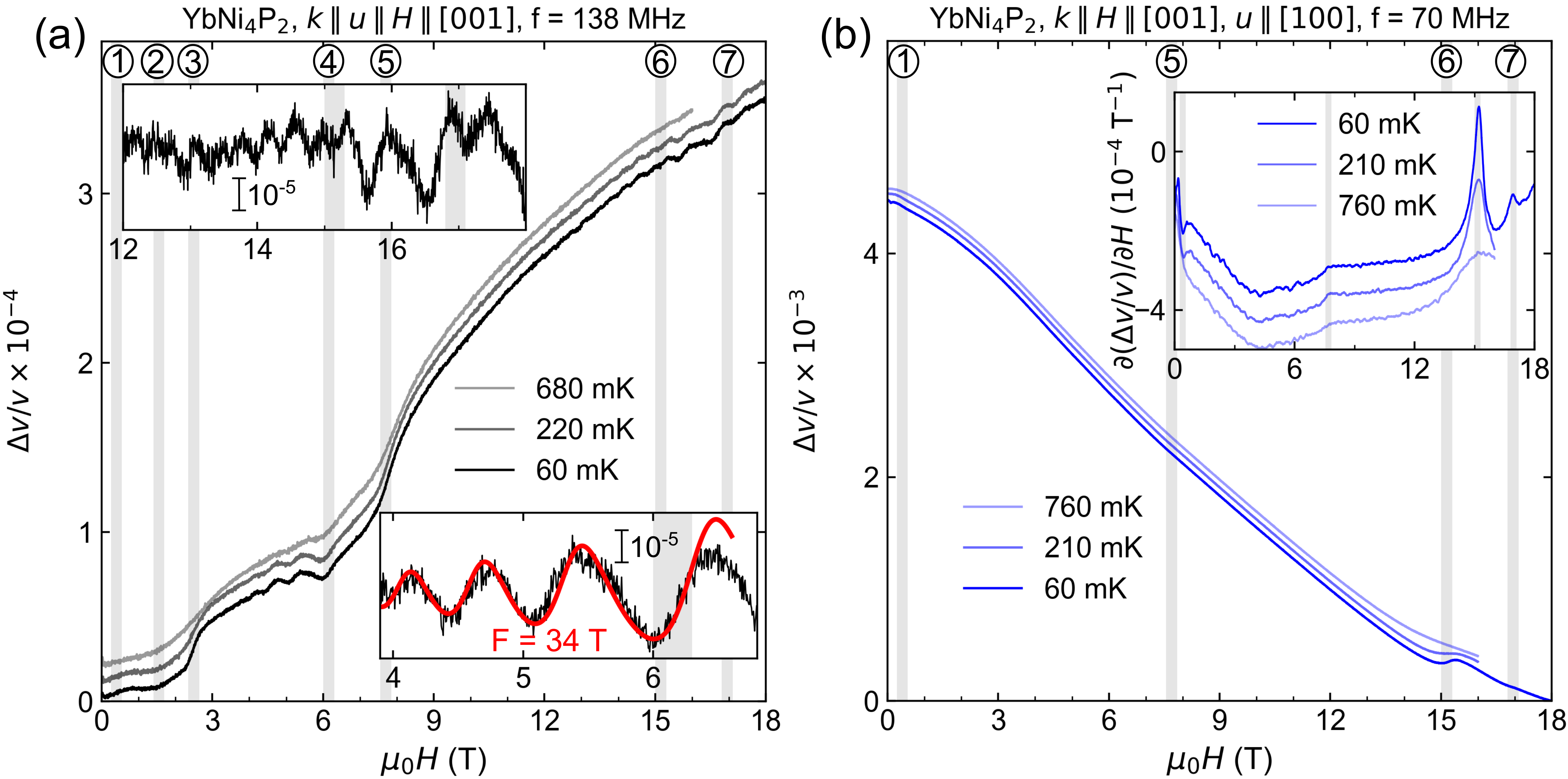}
    \caption{(a) Field-dependent sound-velocity changes for the longitudinal $\mathbf{k}\parallel\mathbf{u}\parallel[001]$ acoustic mode at different temperatures. The insets show quantum oscillations after subtracted background contributions. For the quantum oscillations at low fields, the red line shows a fit using the Lifshitz-Kosevich formula with a Dingle temperature of 0.4 K. (b) Field-dependent sound-velocity changes for the transverse $\mathbf{k}\parallel[001]$, $\mathbf{u}\parallel[100]$ mode at different temperatures. The inset shows the derivative of the sound-velocity change with respect to the field. The curves are arbitrarily shifted for clarity. Locations of the ETT are marked by vertical grey lines.}\label{fig3} 
\end{figure*}

We applied magnetic fields up to 18 T along the [001] axis using a superconducting magnet. We performed ultrasound measurements utilizing the transmission pulse-echo technique with phase-sensitive detection as described in Refs. \onlinecite{luthi2007physical,hauspurg2024fractionalized}. We have set the ultrasound propagation and polarization directions $\textbf{k}$ and $\textbf{u}$ along $[001]$, which corresponds to the direction of the quasi-1D Ni chains, and also in the direction $[100]$, perpendicular to the chains. We attached LiNbO$_3$ transducers (36°-Y cut and 41°-X cut for exciting longitudinal and transverse modes, respectively) to the polished surfaces of the specimen. In this study, we used ultrasound frequencies between 60 and 160 MHz.

\section{Experimental Results}\label{sec3}

We measured the sound-velocity changes as a function of magnetic field for different temperatures. We observed several anomalies, whose positions in field are consistent with the ETT reported previously for this compound \cite{pfau2017cascade}. Remarkably, the different ETT are better detected by particular acoustic modes. We, thus, present the results separately for each mode and summarize the locations of the different anomalies for each elastic mode in Table \ref{USano}.

\begin{table}[b!]
    \centering
     \caption{Anomalies observed for various elastic modes.}
     \label{USano}
     \scalebox{1}{\begin{tabular}{cccccccccccccccc} 
        \toprule%
        \midrule
      Field (T) & & 0.4 & & 1.5 & & 2.5 & & 6.2 & & 7.7 & & 15.2 & & 17.2 \\
      \midrule
      ETT index & & \circled{1} & & \circled{2} & & \circled{3} & & \circled{4} & & \circled{5} & & \circled{6} & & \circled{7} \\
      \midrule
      $\mathbf{k}\parallel\mathbf{u}\parallel[001]$ & & $\times$ & & $\times$ & &  $\times$ & & $\times$ & & $\times$ & & $\times$ & & $\times$  \\
      $\mathbf{k}\parallel[001], \mathbf{u}\parallel[100]$ & & $\times$ & & &  & & & & & $\times$ & & $\times$ & & $\times$ \\
      $\mathbf{k}\parallel\mathbf{u}\parallel[100]$&  &$\times$ & & $\times$ & & $\times$ & & & & $\times$ & & $\times$ & & $\times$  \\
      $\mathbf{k}\parallel[100], \mathbf{u}\parallel[010]$& &$\times$ & &  $\times$ & & &  & $\times$ & & $\times$ & & $\times$ & &  \\
      \midrule
      \bottomrule
     \end{tabular}}
\end{table}

\subsection{Longitudinal acoustic mode $\mathbf{k}\parallel\mathbf{u}\parallel[001]$}

\begin{figure*}
    \centering
    \includegraphics[width=\linewidth]{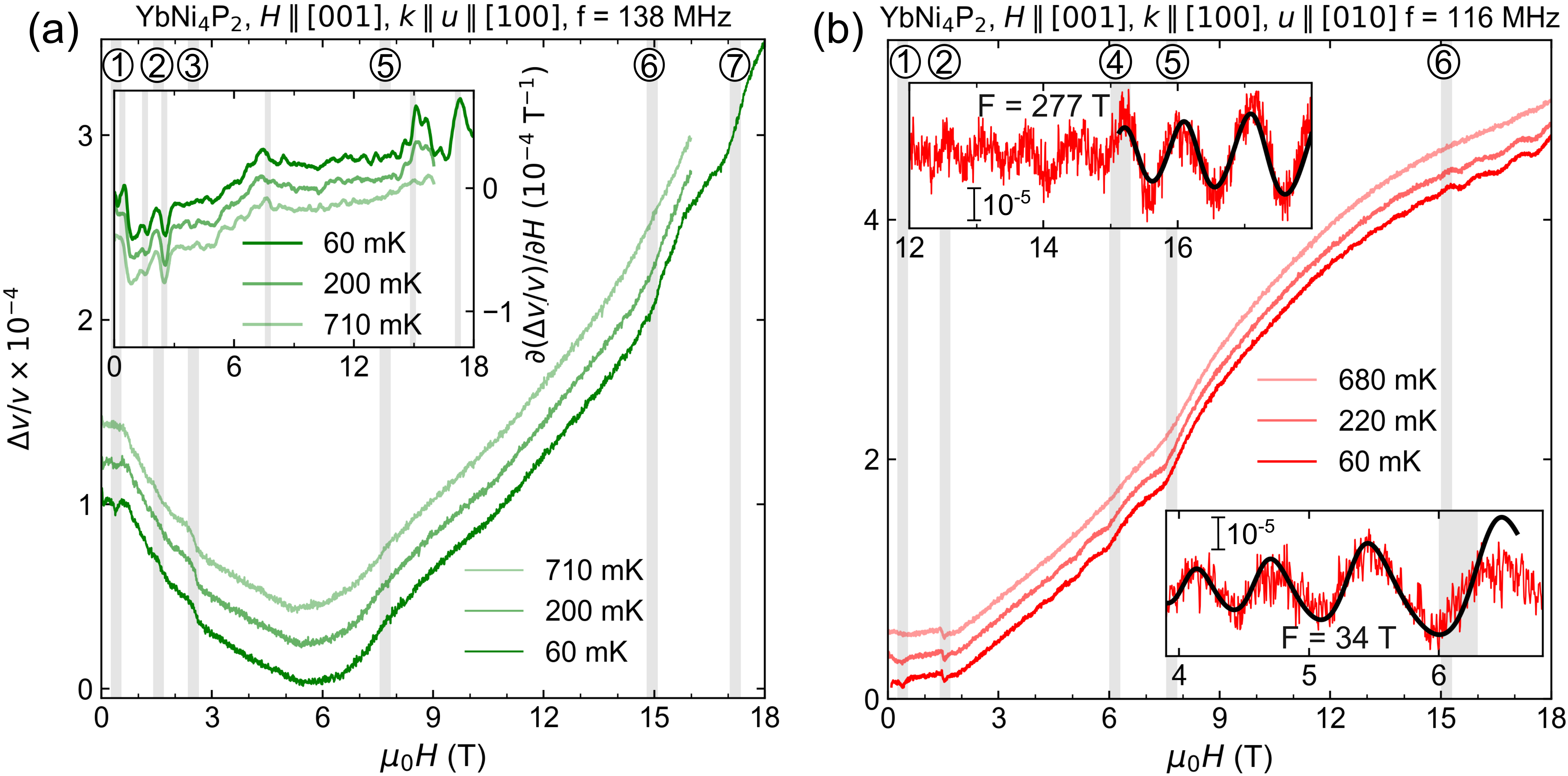}
    \caption{(a) Field-dependent sound-velocity changes for the longitudinal $\mathbf{k}\parallel\mathbf{u}\parallel[100]$ acoustic mode at different temperatures. The inset shows the derivative of the sound velocity change with respect to the field. (b)  Field-dependent sound-velocity changes for the transverse $\mathbf{k}\parallel[100]$, $\mathbf{u}\parallel[010]$ acoustic mode at different temperatures. The insets show quantum oscillations after subtracted background contribution. We performed a fit from the Lifshitz-Kosevich formula with a Dingle temperature of 0.4 K, described in the following section. The curves are arbitrarily shifted for clarity. Locations of the ETT are marked by vertical grey lines. }\label{fig4} 
\end{figure*}

We display our results for the longitudinal ($\mathbf{k}\parallel\mathbf{u}\parallel[001]$) acoustic mode in Fig. \hyperref[fig3]{\ref*{fig3}(a)}. The magnetic field leads to an overall hardening of this mode (increase of $v$ with field). We identify several ETT as changes of slope in the sound velocity, at 0.4, 1.5, 2.5, 6.2, and 7.7 T. Furthermore, the transition at about 6.2 T discloses the end of magnetoacoustic quantum oscillations with a frequency of 34 T. This frequency is very low and indicates a small Fermi-surface orbit disappearing at 6.2 T. We estimate the associated effective mass as $m^* = 1.5$ $m_e$, by analyzing the temperature dependence of the fast-Fourier-transformed (FFT) amplitude. We show this analysis in the Supplemental Material \cite{supplemntary}.

\

Above 12 T, we detect another range of quantum oscillations, with much higher frequencies of the order of 300 T, as shown in the upper inset of Fig. \hyperref[fig3]{\ref*{fig3}(a)}. However, it is more challenging to analyze these high-field oscillations. A possible explanation could be the presence of additional ETT in this field range, in particular at 15.2 T, above which the amplitude of the quantum oscillations increases abruptly, and at 17.2 T, where it decreases abruptly. Thus, the superposition of quantum oscillations and additional sharp anomalies due to ETT could then explain the complex features observed for this acoustic mode at higher fields. We discuss details of the quantum oscillations in the following section.

\subsection{Transverse acoustic mode $\mathbf{k}\parallel[001], \mathbf{u}\parallel[100]$ }

We show our results for the transverse ($\mathbf{k}\parallel[001]$, $\mathbf{u}\parallel[100]$) acoustic mode in Fig. \hyperref[fig3]{\ref*{fig3}(b)}. The field induces an overall softening of the elastic mode with a strong change of the sound velocity, more than one order of magnitude larger compared to the longitudinal mode with $\mathbf{k}\parallel\mathbf{u}\parallel[001]$. Moreover, we observe a distinct set of anomalies for this mode. From the derivatives with respect to the magnetic field, shown in the inset, we detect anomalies at 0.4, 7.7, 15.2, and 17.2 T. We do not observe any quantum oscillation for this mode. Compared to the longitudinal mode with $\mathbf{k}\parallel\mathbf{u}\parallel[001]$, the anomaly at 7.7 T is much less pronounced here, leading only to a kink in the sound-velocity derivative [inset of Fig. \hyperref[fig3]{\ref*{fig3}(b)}]. On the other hand, the anomalies at 15.2 and 17.2 T are much sharper.

\subsection{Longitudinal acoustic mode $\mathbf{k}\parallel\mathbf{u}\parallel[100]$}

We show our results for the longitudinal ($\mathbf{k}\parallel\mathbf{u}\parallel[100]$) acoustic mode in Fig. \hyperref[fig4]{\ref*{fig4}(a)}. For this mode, the magnetic field induces, first, a softening up to 6 T and then a hardening. The total variation of the sound velocity is comparable to the longitudinal mode propagating along $[001]$. We do not observe any quantum oscillation. As for the previous elastic modes, we detect several anomalies in the sound velocity, which are better visible in the sound-velocity derivatives in the inset of Fig. \hyperref[fig4]{\ref*{fig4}(a)}. We observe clear features at 0.4, 1.5, 2.5, 7.7, 15.2, and 17.2 T.

\subsection{Transverse acoustic mode $\mathbf{k}\parallel[100], \mathbf{u}\parallel[010]$ }

We show our results for the transverse ($\mathbf{k}\parallel[100]$, $\mathbf{u}\parallel[010]$) acoustic mode in Fig. \hyperref[fig4]{\ref*{fig4}(b)}. Remarkably, the curves are very similar to our results for the longitudinal mode with $\mathbf{k}\parallel\mathbf{u}\parallel[001]$. In particular, we observe quantum oscillations below 6 T as well as above 12 T.   

Our analysis of the quantum oscillations below 6.2 T leads to the same frequency of 34 T as extracted for the longitudinal mode with $\mathbf{k}\parallel\mathbf{u}\parallel[001]$. However, above 12 T, our results reveal new features. In Fig. \hyperref[fig4]{\ref*{fig4}(b)}, only one frequency of 277 T fits well the data between 15 and 18 T, while for the data from Fig. \hyperref[fig3]{\ref*{fig3}(a)} the fit is less clear (not shown). This might indicate that the transverse acoustic mode with $\mathbf{k}\parallel[100]$, $\mathbf{u}\parallel[010]$ is less sensitive to the ETT at 17.2 T. In contrast, we clearly detect the ETT at 15.2 T for both modes, with an abrupt change of the quantum-oscillation amplitude.

\section{Analysis of the quantum oscillations}

In this section, we analyze the quantum oscillations detected for the longitudinal $\mathbf{k}\parallel \mathbf{u} \parallel [001]$ and the transverse $\mathbf{k}\parallel [100]$, $\mathbf{u} \parallel [010]$ acoustic modes. We obtained the effective masses $m^*/m_e$ by analyzing the temperature dependence of the FFT amplitudes, shown in the Supplemental Materials \cite{supplemntary}. We evaluated the effective masses only for the longitudinal mode with $\mathbf{k}\parallel \mathbf{u} \parallel [001]$, due to the limited number of available temperatures for the transverse mode with $\mathbf{k}\parallel [100]$, $\mathbf{u} \parallel [010]$. Furthermore, we obtained the mean free path $l_0$ and the Dingle temperature $T_D$ by analyzing the field dependence of the FFT amplitudes at the lowest temperature. Our main results are summarized in Table \ref{QuantoOsc}.

\begin{table}[b!]
    \centering
     \caption{Summary of quantum-oscillation frequencies and cyclotron masses. Effective masses $m^*$ are extracted by fitting the Lifshitz-Kosevich temperature dependence to the FFT peak amplitudes. Mean-free paths $l_0$ are extracted from the field dependence to the FFT peak amplitudes.}
     \label{QuantoOsc}
     \scalebox{1}{\begin{tabular}{ccccccccc} 
        \toprule%
        \midrule
      Acoustic mode & & $F$(T) & & $m^*/m_e$ & & $l_0$ (nm) & & $T_D$ (K) \\
      \midrule
      $\mathbf{k}\parallel\mathbf{u}\parallel[001]$ & & 34 & & 1.5 & &  80(23) & & 0.37(9)  \\
      $\mathbf{k}\parallel\mathbf{u}\parallel[001]$ & & 125 & & 5 & &  - & & -  \\
      $\mathbf{k}\parallel\mathbf{u}\parallel[001]$ & & 310 & & 6.5 & &  14(2) & & 1.5(2)  \\
      $\mathbf{k}\parallel\mathbf{u}\parallel[001]$ & & 535 & & 3 & &  153(26) & & 0.34(3)  \\
      \midrule
      $\mathbf{k}\parallel [100]$, $\mathbf{u} \parallel [010]$ & & 34 & & - & &  - & & -  \\
      $\mathbf{k}\parallel [100]$, $\mathbf{u} \parallel [010]$ & & 277 & & - & &  - & & -  \\
      \midrule
      \bottomrule
     \end{tabular}}
\end{table}

We remark on the presence of relatively large effective masses for the frequencies of 125 and 310 T, in agreement with the heavy-fermion character of YbNi$_4$P$_2$. Furthermore, the estimated temperatures for the small pocket of 34 T and the large pocket at 535 T result in a value of around 0.4 K. We, thus, propose a fit of the quantum oscillations for both longitudinal and transverse modes with the Lifshitz-Kosevich formula, which for the sound velocity gives \cite{schindler2020strong,shoenberg2009magnetic}:
\begin{equation}\label{0}
  \begin{split}
    \frac{\Delta v}{v} &= A_0 \sum_{p=1}^\infty p^{-\frac{1}{2}}R_TR_D\cos\left[2\pi p \left(\frac{F}{B} - \phi\right) \pm \frac{\pi}{4}\right],\\
    R_T & = \frac{\lambda(T)}{\sinh[\lambda(T)]}, \ \ \ R_D = \exp[-\lambda(T_D)], \\ \lambda(T) &= p \frac{2\pi^2 m^* k_B T}{e \hbar H},
  \end{split}
\end{equation}
where $A_0$, $\phi$ and $F$ were adjusted to describe the data in Fig. \hyperref[fig3]{\ref*{fig3}(a)} and \hyperref[fig4]{\ref*{fig4}(b)}. For the 34 T oscillations, we used a phase shift of $+\pi/4$, while for the 277 T oscillations we cannot extract a reliable phase, probably due to the higher effective mass. We assume that the orbit of 277 T, observed for the mode with $\mathbf{k}\parallel [100]$, $\mathbf{u} \parallel [010]$, corresponds to the orbit of 310 T observed for the mode with $\mathbf{k}\parallel \mathbf{u} \parallel [001]$, where the small difference could arise from small misalignments of the sample between the two experiments.

\begin{figure}[t!]
    \centering
    \includegraphics[width=\linewidth]{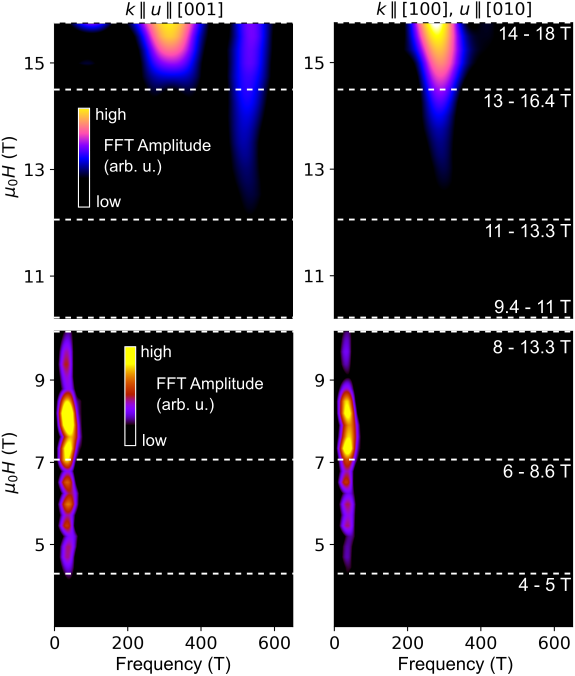}
    \caption{Contour plot of the field dependence of the FFT amplitudes. In the upper panels, the FFT was performed at a constant inverse-field window of $1/14 - 1/18 = 0.0159$ T$^{-1}$, while in the lower panels the window was set to $1/4 - 1/5 = 0.05$ T$^{-1}$. The value of $\mu_0H$ on the laft axis was determined from the harmonic mean of the FFT window. Step width was set at 0.25 T.}\label{fig5} 
\end{figure}

We show the field dependence of the FFT amplitudes of both the longitudinal $\mathbf{k}\parallel \mathbf{u} \parallel [001]$ and the transverse $\mathbf{k}\parallel [100]$, $\mathbf{u} \parallel [010]$ acoustic modes in Fig. \hyperref[fig5]{\ref*{fig5}}. At low fields, we observe the appearance of the peak of frequency 34 T in both acoustic modes for magnetic fields above 4 T. This peak is clearly visible until around 9 T and disappears at higher fields. From the frequency of 34 T, one can estimate a quantum-limit field $B_Q \approx F \approx 34$ T. Thus, at 9 T, we are well below the quantum limit, and the disappearance of this frequency directly evidences an electronic-topological transition.

At higher fields, we observe the appearance of new frequencies. For the $\mathbf{k}\parallel \mathbf{u} \parallel [001]$ mode, we can resolve three frequencies at $\mu_0H=$ 15.75 T, which corresponds to the field window between 14 and 18 T. In particular, the frequency of 125 T is visible only at the highest fields, which explains why we cannot perform the analysis of the field dependence of the FFT amplitude in this case. While the oscillations of 310 T are visible over a broader field range, this is still not sufficient to obtain a good fit. Finally, the frequency of 535 T is visible over a large field range between 12 and 18 T, which allows us to obtain a quite reliable value for the mean-free path in Table \ref{QuantoOsc}.

We remark on the strong difference between the case of the longitudinal mode with $\mathbf{k}\parallel \mathbf{u} \parallel [001]$, where we identify three frequencies at highest fields, and the case of the transverse mode with $\mathbf{k}\parallel [100]$, $\mathbf{u} \parallel [010]$, where we obtain a single frequency. We propose that this could arise from a smaller coupling between the transverse mode and the 125 and 535 T orbits.

\

Finally, we emphasize the presence of potential artifacts in our analysis of the magneto-acoustic quantum oscillations arising from the non-trivial background subtraction in the presence of the ETT leading to sharp features in the raw data. We expect that such artifacts could be responsible for the presence of the peak at 34 T in the FFT amplitude up to 9 T in Fig. \hyperref[fig5]{\ref*{fig5}}, while the raw data in Fig. \hyperref[fig3]{\ref*{fig3}(a)} show the absence of quantum oscillations between 6 and 9 T. Furthermore, since both the ETT and the quantum oscillations arise from the same charge carriers, we cannot expect to distinguish between them in the temperature dependence, for example. While these artifacts might not affect our interpretation for the 34 T and 277 T orbits, where the Lifshitz-Kosevich formula fits the experimental data, our results for the other orbits are less certain. In particular, the case of the 125 T orbit needs further investigations at higher magnetic fields.

\section{Theory}

In order to discuss in more detail our observations of electronic-topological transitions and magnetoacoustic quantum oscillations in YbNi$_4$P$_2$, we developed a detailed analysis of the electronic structure in the presence of strong electronic correlations and magnetic fields, as well as a toy model for the electron-phonon coupling. We define our total Hamiltonian as
\begin{equation}\label{1}
  \begin{split}
    \mathcal{H} = \mathcal{H}_e + \mathcal{H}_{ph} + \mathcal{H}_{e-ph},
  \end{split}
\end{equation}
where $\mathcal{H}_e$ describes the electronic degrees of freedom, $\mathcal{H}_{ph}$ describes the lattice, and $\mathcal{H}_{e-ph}$ is an electron-phonon interaction term, defined hereafter.

\subsection{Electronic structure and Fermi surfaces in field}

\begin{figure}
    \centering
    \includegraphics[width=\linewidth]{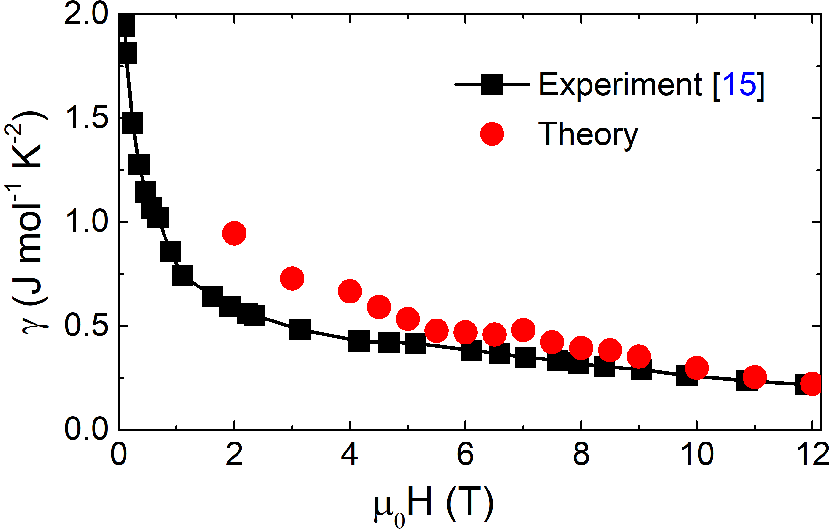}
    \caption{Magnetic-field dependence of the electronic Sommerfeld coefficient (experimental points from \cite{pfau2017cascade}).}\label{fig0b} 
\end{figure}

We carried out our electronic-structure calculations using the renormalized band method (RB) \cite{zwicknagl2011field,zwicknagl2016utility}, which allows a precise investigation of the field-induced ETT in heavy-fermion systems \cite{pourret2019transport}. This method contains an ab-initio description of the non-correlated electron bands thanks to the fully relativistic formulation of the linear muffin-tin orbitals method \cite{andersen1975linear, skriver2012lmto, albers1986electronic}. It also incorporates the precise crystal-electric-field scheme for the Yb ions and phenomenological considerations for the strong electronic correlations of the Yb $f$ states in the spirit of Fermi-liquid theory \cite{zwicknagl2016utility}. At zero field, RB reduces to a treatment of the correlated Yb $4f$ holes with the slave-boson mean-field theory \cite{barnes1976new,coleman1987mixed,akbari2009theory,burdin2009multiple,riseborough2016mixed,sourd2024nonlocal}. At non-zero magnetic fields the RB scheme includes the progressive de-renormalization of the quasiparticles as well as a correlation-enhanced Zeeman splitting obtained from the renormalized perturbation theory of the single-impurity Anderson model \cite{hewson2006field}. Details of our calculations are described in the Supplemental Materials \cite{supplemntary}. We neglect the contribution of the ferromagnetic order, since the magnetic moment of 0.05 $\mu_B$ per Yb ion is very small \cite{krellner2011ferromagnetic}. We used the crystal-electric-field states and splittings determined previously for YbNi$_4$P$_2$ \cite{huesges2018analysis}, and we adjusted the phenomenological parameter to reproduce the Sommerfeld coefficient $\gamma = 1.5$ Jmol$^{-1}$K$^{-2}$ reported from specific-heat measurements in the polarized state at 200 mT \cite{krellner2011ferromagnetic}. 

\begin{figure*}
    \centering
    \includegraphics[width=\linewidth]{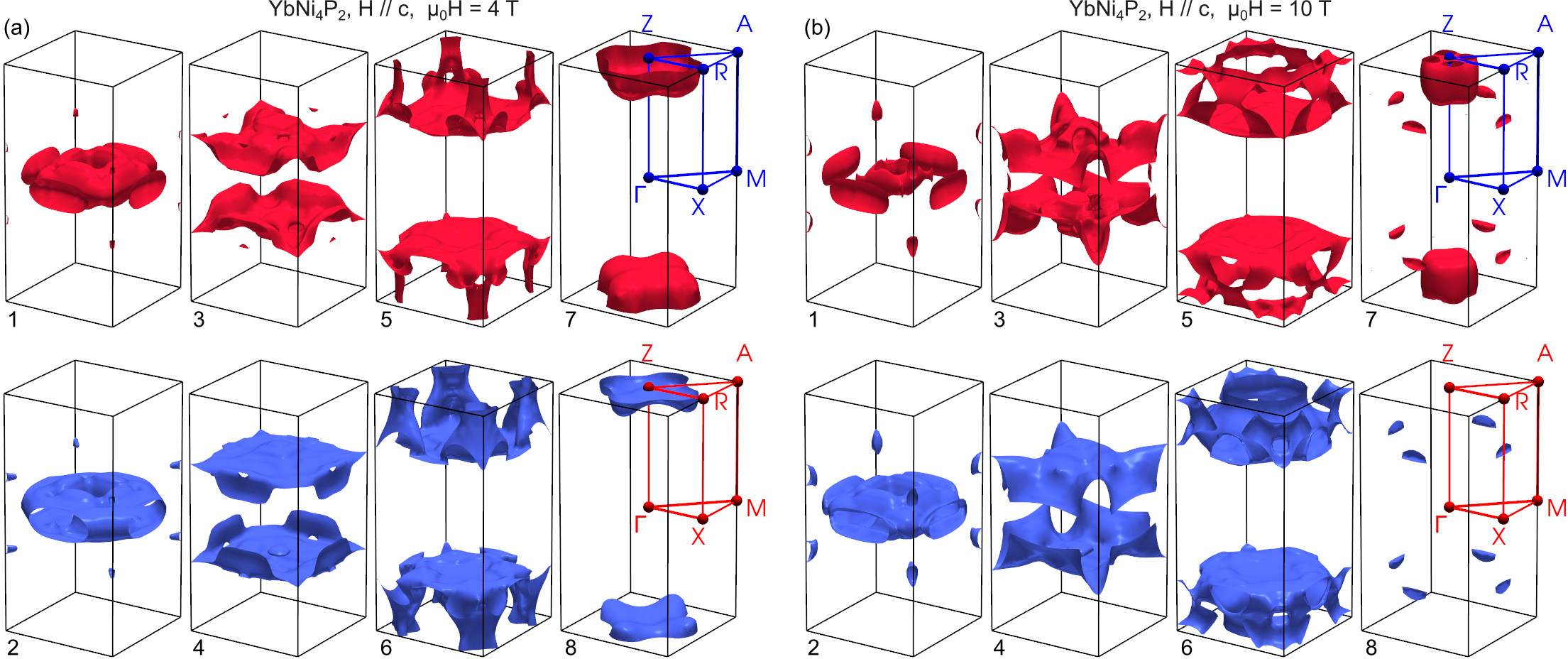}
    \caption{Fermi surfaces at a magnetic field (a) of 4 T and (b) 10 T. The Fermi surfaces for bands 1, 3, 5, and 7 correspond to the majority spin, while the Fermi surfaces for bands 2, 4, 6, and 8 correspond to the minority spin.}\label{fig1b} 
\end{figure*}

In order to evaluate the electron-phonon coupling later, it is useful to express the RB result as an effective tight-binding Hamiltonian $\mathcal{H}_e$. $\mathcal{H}_e$ contains conduction states $c_{q\eta\sigma}$ of quasi-momentum $q$, orbital index $\eta$, and spin $\sigma$ hybridized with renormalized $\tilde{f}_{q\gamma\sigma}$ pseudofermion states of spin-orbital index $\gamma$ and pseudospin $\sigma$. We write it as
\begin{equation}\label{Eq1}
    \begin{split}
 \mathcal{H}_e &=  \sum_{q\eta\eta'\sigma} \left[\epsilon_q^{\eta\eta'}-\left(\mu + \sigma \frac{g\mu_B h}{2} \right)\delta^{\eta\eta'}\right] c_{q\eta\sigma}^\dagger c_{q\eta'\sigma} \\
  &+ \sum_{q\eta\gamma\sigma}\left( b(h)\epsilon_q^{\eta\gamma} c_{q\eta\sigma}^\dagger \tilde{f}_{q\gamma\sigma} + \text{H.c.}\right) \\ 
  &+ \sum_{q\gamma\sigma}\left(\epsilon_{f}+\Delta_\gamma(h)+\lambda \right)\tilde{f}_{q\gamma\sigma}^{\dagger}\tilde{f}_{q\gamma\sigma} - \lambda N_f + \mu N_c
    \end{split}{}
\end{equation}{}
where $\epsilon_{q}^{\eta\eta'}$ are the conduction-hole energies, $\mu$ is the chemical potential, $h$ is the external magnetic field, $N_f$ and $N_c$ are the number of $f$ and conduction holes, respectively, $\epsilon_{q}^{\eta\gamma}$ the hybridization between conduction and $f$ states, $\epsilon_f$ and $\Delta_f(h)$ the on-site energy and crystal field splittings of the $f$ states, respectively, and $b(h)$ and $\lambda$ correspond to the slave-boson mean-field parameters in the limit $h\rightarrow 0$. For non-zero magnetic field, the progressive de-renormalization of the quasiparticles is encoded in the variation of $b(h)$, while the enhanced Zeeman splittings are encoded in $\Delta_f(h)$. The Fermi-liquid assumption of the RB scheme corresponds to a constant $N_f$ for all values of the magnetic field. This constraint is enforced on average by the Lagrange multiplier $\lambda$. For a compact notation, we write this Hamiltonian in matrix form as
\begin{equation}\label{Eq3}
    \begin{split}
 \mathcal{H}_{e} &= \sum_{q\sigma} \boldsymbol{\Psi}_{q\sigma}^\dagger \bar{\mathcal{H}}_q \boldsymbol{\Psi}_{q\sigma} - \lambda N_f + \mu N_c,\\
  \mathcal{H}_q^{\alpha\alpha'} &= \mu^{\alpha\alpha'}(h) + \epsilon_q^{\alpha\alpha'}, \\
  \epsilon_q^{\alpha\alpha'} &=\sum_\delta  t_\delta^{\alpha\alpha'} e^{i\mathbf{q}\cdot\boldsymbol{\delta}}\left(\delta^{\alpha\eta}\delta^{\alpha'\eta} + b(h)\delta^{\alpha\eta}\delta^{\alpha'\gamma}\right),
    \end{split}{}
\end{equation}{}

where  $\boldsymbol{\Psi}_{q\sigma}^\dagger = (c_{q\eta_0\sigma}^\dagger,c_{q\eta_1\sigma}^\dagger,...\tilde{f}_{q\gamma_0\sigma}^\dagger,..,\tilde{f}_{q\gamma_7\sigma}^\dagger)$ is a vector of creation operators indexed by the new orbital index $\alpha$, $t_\delta^{\alpha\alpha'}$ are the effective tight-binding hopping parameters, and $\boldsymbol{\delta}$ denotes the connecting vector between any pair of sites. From the density of states at the Fermi level for different values of the magnetic field, we obtained the field dependence of the Sommerfeld coefficient shown in Fig. \hyperref[fig0b]{\ref*{fig0b}}. This result is in good agreement with the experimental data reported in \cite{pfau2017cascade}, particularly above 4 T. We obtained four bands crossing the Fermi level, leading to strongly anisotropic Fermi surfaces as shown in Figs. \hyperref[fig1b]{\ref*{fig1b}(a)} and \hyperref[fig1b]{\ref*{fig1b}(b)}, for magnetic fields of 4 and 10 T respectively.

Our predictions agree well with very recent ARPES results \cite{dai2025electronic}, namely 1D sheets in the $\Gamma \rightarrow Z$ path ($q_x = q_y = 0$, $q_z = k$) for bands 3, 4, 5 and 6, and mixed dimensionality sheets on the $q_z = 0$ (bands 1 and 2) and $q_z = \pi/c$ (bands 7 and 8). Furthermore, by comparing the Fermi surfaces between 4 and 10 T, we can identify several electronic-topological transitions (ETT), particularly visible on the $q_z = \pi/c$ plane. For bands 5 and 6, we observe the evolution from 4 ``tube-like" Fermi surfaces around the corner of the Brillouin zone at low field, which merge into a large circle around the $Z$ point for higher fields. For bands 7 and 8, the presence of ETT is even more clear, with the complete disappearance of band 8 at high fields, which is replaced by four smaller pockets. In the following, we show how a detailed modeling of our ultrasound results can allow us to investigate further these transitions.

\subsection{Phonons $\mathcal{H}_{ph}$ and electron-phonon coupling $\mathcal{H}_{e-ph}$}
During the ultrasound experiment, we produce an external perturbation of the form of a displacement wave $\mathbf{u}(\mathbf{r},t)= \mathbf{u}_k\exp(i\mathbf{k}\cdot\mathbf{r}-i\omega t)$, where $\mathbf{u}_k$ and $\mathbf{k}$ are, respectively, the sound-wave polarization and propagation vectors. For frequencies of the order of 100 MHz, we are typically in the regime $\mathbf{k}\cdot\boldsymbol{\delta} \ll 1$, where $|\boldsymbol{\delta}|$ is of the order of the lattice spacing. Thus, we will be interested in the limit $\mathbf{k}\rightarrow 0$ in our calculations. In order to evaluate the effect of such perturbation on the electron Hamiltonian $\mathcal{H}_e$ of Eq. (\ref{Eq3}), we write the position of a given atom as $\mathbf{R}_i = \mathbf{R}_i^0 + \mathbf{Q}_i$ with the equilibrium position $\mathbf{R}_i^ 0$ and the displacement $\mathbf{Q}_i$. We then quantize the atomic displacement $\mathbf{Q}_i$ with the usual phonon annihilation and creation operators $a_{k}$ and $a_{k}^\dagger$, respectively \cite{mahan2000many}:
\begin{equation}\label{eq9}
    \begin{split}
        \mathbf{Q}_i &= i \sum_{k} \sqrt{\frac{1}{2M_0N\omega_{k}^0}} \mathbf{u}_{k} (a_{k}+a_{-k}^\dagger)e^{i\mathbf{k}\cdot\mathbf{R}_i^0},
    \end{split}{}
\end{equation}{}
where $M_0$ denotes the mass of a YbNi$_4$P$_2$ unit cell, $N$ the number of unit cells, and $\omega_{k}^0$ the energy of the phonon excitation. Thus, the informations that permit distinguishing each acoustic mode are the direction of $\mathbf{u}_k$ and $\mathbf{k}$.

As described in detail in the Supplemental Materials \cite{supplemntary}, we consider three different contributions to the electron-phonon couplings. The first one arises from the Taylor expansion of the effective tight-binding parameter $t_{\delta}^{\alpha\alpha'}$ upon varying the bond vector $\boldsymbol{\delta}$ due to the displacement $\mathbf{Q}_{i} - \mathbf{Q}_{i+\delta}$ \cite{barivsic1972rigid, walker2001electron}. We use a simple ansatz of an exponential decrease with increasing bond distance $t_{\delta}^{\alpha\alpha'}  = (t_{\delta}^{\alpha\alpha'})^0 \text{exp}(- \xi_0 |\mathbf{R}_{i}-\mathbf{R}_{i+\delta}| + \xi_0\delta)$, where we suppose the typical inverse length scale $\xi$ to be the same for each atomic overlap. With this choice, the first-order term of the Taylor expansion is easily evaluated:
\begin{equation}\label{eq12}
    \begin{split}
        \sum_\mu u_k^\mu \frac{\partial t_{\delta}^{\alpha\alpha'} }{\partial R^\mu} &= -\xi_0 (t_{\delta}^{\alpha\alpha'})^0 (\boldsymbol{\delta} \cdot \mathbf{u}_k),
    \end{split}{}
\end{equation}{}
where $\mathbf{u}_k$ is the phonon polarization. From the factor $(\boldsymbol{\delta} \cdot \mathbf{u}_k)$, we see that only the atomic displacements stretching or compressing the chemical bonds contribute. The two other contributions to the electron-phonon coupling are specific to the heavy-fermion physics and arise from variations of the slave-boson mean-field parameters $b(h)$ and $\lambda$ with strain, as discussed in Refs. \onlinecite{fulde1988theory,keller1990electron}. They originate from the strong pressure dependence of the Kondo temperature and the associated enhancement of the Grüneisen parameter in heavy-fermion systems \cite{fulde1988theory}. In order to incorporate the microscopic displacements of each Yb, Ni, and P atom in this scheme, we use the fact that $b(h)$ and $\lambda$ arise mainly from the non-local Kondo coupling between the Yb $f$ orbitals and the Ni $d$ (P $p$) orbitals \cite{sourd2024nonlocal}. The non-local Kondo coupling is taken care of by assuming bond-dependent terms $b_\delta^\gamma = b(h)\beta_\delta^\gamma $ and $\lambda_\delta^\gamma = \lambda\beta_\delta^\gamma$ such that
\begin{equation}\label{eq10}
    \begin{split}
        \beta_\delta^\gamma =  \frac{\sum_{\eta} |t_{\delta}^{\eta \gamma}|}{\sum_{\eta\delta} |t_{\delta}^{\eta \gamma}|} \Rightarrow \sum_{\delta} \beta_\delta^\gamma = 1.
    \end{split}{}
\end{equation}{}
$\beta_\delta^\gamma$ reflects how important is the contribution of the bond $\boldsymbol{\delta}$ in the value of $b(h)$ and $\lambda$. It allows us to write a Taylor expansion of the mean-field parameters $b_\delta^\gamma$ and $\lambda_\delta^\gamma$ upon variation of the bond vector $\boldsymbol{\delta}$ due to the displacement $\mathbf{Q}_{i} - \mathbf{Q}_{i+\delta}$, where we assume the following derivatives
\begin{equation}\label{eq12}
    \begin{split}
         \sum_\mu u_k^\mu \frac{\partial b_{\delta}^\gamma}{\partial R^\mu} &= - \frac{\tilde{\eta} \xi_0 b^0}{2}\beta_\delta^\gamma (\boldsymbol{\delta} \cdot \mathbf{u}_k), \\  \sum_\mu u_k^\mu \frac{\partial \lambda_{\delta}^\gamma}{\partial R^\mu} &= - \tilde{\eta} T_K \xi_0 \beta_\delta^\gamma (\boldsymbol{\delta} \cdot \mathbf{u}_k),
    \end{split}{}
\end{equation}{}
where $T_K = 8$ K is the Kondo temperature of YbNi$_4$P$_2$. These two terms are proportional to the dimensionless parameter $\tilde{\eta}$, which accounts for the enhanced Grüneisen parameter in the heavy-fermion regime \cite{fulde1988theory,keller1990electron}. For YbNi$_4$P$_2$, we obtaine $\tilde{\eta} \approx 4.5$ at 5 T. Assuming the phonon dynamics to be harmonic with energy $\omega_k^0$, we obtain the lattice Hamiltonian and the electron-phonon coupling as:
\begin{equation}\label{eq10}
    \begin{split}
       \mathcal{H}_{ph} &= \sum_{k} \omega_{k}^0 a_{k}^\dagger a_{k},\\
            \mathcal{H}_{e-ph} &= \sum_{kq\sigma} \boldsymbol{\Psi}_{q+k\sigma}^\dagger \bar{M}_{qk} \boldsymbol{\Psi}_{q\sigma}\left(a_k + a_{-k}^\dagger\right), \\
            M_{qk}^{\alpha\alpha'} &= -\frac{i\xi}{\sqrt{2\omega_k^0}}\left[ \sum_{\delta} t_{\delta}^{\alpha\alpha'} e^{i\mathbf{q} \cdot \boldsymbol{\delta}}\left(M_t + M_b \right) + M_\lambda\right], \\
            M_t & = \delta^{\alpha\eta}\delta^{\alpha'\eta}\left(1-e^{i\mathbf{k} \cdot \boldsymbol{\delta}}\right)(\boldsymbol{\delta}\cdot \mathbf{u}_k),\\
            M_b &= \frac{b\tilde{\eta}}{2}  \delta^{\alpha\eta}\delta^{\alpha'\gamma}\sum_{\delta'}\beta_{\delta'}^\gamma(\boldsymbol{\delta}'\cdot \mathbf{u}_{k})\left(1-e^{i\mathbf{k} \cdot \boldsymbol{\delta}'}\right), \\
            M_\lambda &= T_K\tilde{\eta}\delta^{\alpha\gamma}\delta^{\alpha'\gamma} \sum_{\delta'}\beta_{\delta'}^\gamma(\boldsymbol{\delta}'\cdot \mathbf{u}_{k})\left(1-e^{i\mathbf{k} \cdot \boldsymbol{\delta}'}\right),
    \end{split}{}
\end{equation}{}
with $\xi = \xi_0/\sqrt{M_0}$. In this model, the lattice Hamiltonian $\mathcal{H}_{ph}$ is field independent, and all the variations of the phonon spectrum in field come from the electron-phonon coupling. We obtain the sound-velocity variation with field as the real part of the (field-dependent) phonon self-energy $\Sigma$:
\begin{equation}\label{eq15}
    \begin{split}
   \frac{\Delta v}{v} &= \lim_{k\rightarrow 0} \frac{\Delta \omega_k^0}{\omega_k^0} = \lim_{k\rightarrow 0} \frac{\text{Re} \Sigma(k,i\nu_n)_{i\nu_n \rightarrow \omega_k^0 + i0^+}}{\omega_k^0} + \Delta_0\\
   &= \frac{\xi^2}{v^2}\Sigma_1 + \Delta_0,
    \end{split}{}
\end{equation}{}
where $ \lim_{k\rightarrow 0} \omega_k^0 = vk$,  $\Delta_0$ is an offset and $\Sigma_1$ will be defined in the following section. 
\subsection{Phonon self-energy}
In our ultrasound experiments, the phonon energy is very small (of the order of $\mu$eV), which allows us to use Migdal's theorem and evaluate the phonon self-energy without any vertex correction. We define the electron Green function as $\bar{G}_0(q, \tau) = - \left\langle \mathcal{T}\boldsymbol{\Psi}_q(\tau) \boldsymbol{\Psi}_q^\dagger(0) \right\rangle$, where $\mathcal{T}$ is the chronological-order operator. This function is diagonal in the band basis $\beta$, with  $\bar{G}_0(q, i\nu_n)^{\beta\beta} = (i\nu_n - E_q^\beta+i\text{sgn}(\nu_n)\Gamma)^{-1}$. In this expression, we take $\Gamma$ as an ad-hoc parameter representing scattering on impurities. This term permits avoiding singularities upon evaluating the phonon self-energy at zero temperature, as is used, for example, in the theoretical description of ultrasound experiments on ZrTe$_5$ \cite{ehmcke2021propagation}. We evaluate the phonon self-energy from standard functional-integral techniques \cite{altland2010condensed}. The main contributions to the self-energy and $\Sigma_1$ can be written as

\begin{equation}\label{Eq5}
  \begin{split}
    \Sigma^1(k,i\nu_n)
        &= \sum_{q,i\nu_m} \Tr \left\{\bar{M}_{q,q+k}\bar{G}_0(q,i\nu_m)\times\right. \\
        & \ \ \ \ \ \ \ \ \ \ \ \ \ \ \ \ \ \ \left. \bar{M}_{q+k,-k}\bar{G}_0(q+k,i\nu_m+i\nu_n) \right\},\\
      \Rightarrow \Sigma_1 &= -\frac{1}{\pi}\sum_q\left[\sum_{\beta}|M_q^{\beta\beta}|^2\frac{\Gamma}{(E_q^\beta)^2 + \Gamma^2} \right.\\
      &\left.+ 2\sum_{\beta\neq\beta'}|M_q^{\beta\beta'}|^2\frac{\arctan(\epsilon_\Gamma)-\arctan(\epsilon_\Gamma')}{E_{q}^{\beta'} -E_{q}^{\beta}}\right],\\
      M_{q}^{\beta\beta'} &= \sum_{\alpha\alpha'}M_{q}^{\alpha\alpha'}a_{q}^{\beta\alpha}(a_q^{\alpha'\beta'})^*, \\ 
      M_{q}^{\alpha\alpha'} &= \lim_{k\rightarrow 0} \frac{M_{qk}^{\alpha\alpha'}}{k},
  \end{split}{}
\end{equation}{}
where $\epsilon_\Gamma = \Gamma/ \left[E_{q}^{\beta}+\sqrt{(E_{q}^{\beta})^2 + \Gamma^2}\right]$, $\epsilon_\Gamma' = \Gamma/ \left[E_{q}^{\beta'}+\sqrt{(E_{q}^{\beta'})^2 + \Gamma^2}\right]$, and $a_{q}^{\beta\alpha}$ is the coefficient between the orbital basis $\alpha$ and the band basis $\beta$. The two terms of $\Sigma_1$ can be understood as intraband and interband contributions. For the intraband contribution, we recognize the Dirac delta function of scattering time $\Gamma$: $\delta_\Gamma(\omega) = \Gamma/\pi(\omega^2+\Gamma^2)$. Thus, it can be regarded as the density of states at the Fermi level, multiplied by a factor $|M_{q}^{\beta\beta}|^2$. This term gives the main contribution to the field-induced ETT due to the presence of van Hove singularities, as shown for YbRh$_2$Si$_2$ \cite{naren2013lifshitz}. Furthermore, it is also responsible for the quantum oscillations of the sound velocity, as shown for ZrTe$_5$ \cite{ehmcke2021propagation}.

\

The second term of $\Sigma_1$ is an interband contribution and has a similar prefactor $|M_{q}^{\beta\beta'}|^2$ arising from electron-phonon coupling. It has been proposed in the case of YbRh$_2$Si$_2$ that interband interference effects might be relevant when several ETT occur in a small field window \cite{pourret2019transport}. The situation might be similar in our case. However, if we restrict our study to the quantum oscillations of 34 T, this term can be neglected compared to the first one.

\section{Ultrasound vertices and Discussion}

We thus obtained a potentially singular contribution from the density of states, weighted by the factor $|M_{q}^{\beta\beta}|^2$. In analogy with the Raman study of the ETT in the cuprate superconductor Bi$_2$Sr$_2$CaCu$_2$O$_{8+\delta}$ \cite{benhabib2015collapse}, where the results are discussed in terms of a Raman vertex-weighted density of states, we propose to introduce the notion of an ultrasound vertex in order to discuss our results. The Raman vertex is defined as an electron-photon coupling that possibly vanishes in some regions of the Brillouin zone depending on the light-wave polarization and propagation directions \cite{devereaux2007inelastic}. Thus, in analogy, the ultrasound vertex will correspond to an electron-phonon coupling that possibly vanishes in some regions of the Brillouin zone depending on the acoustic-wave polarization and propagation directions. From the model described above, we define the ultrasound vertex as the factor $|M_{q}^{\beta\beta'}|^2$, which can eventually suppress the potentially singular contribution from the density of states or from the interband interference term, depending on the symmetry of the acoustic mode. In the following, we show that this vertex is a useful tool to discuss the field-induced ETT Nos. 4 and 5.

\subsection{ETT No. 4 at $\mu_0 H \approx 6.2$ T}

\begin{figure*}
    \centering
    \includegraphics[width=\linewidth]{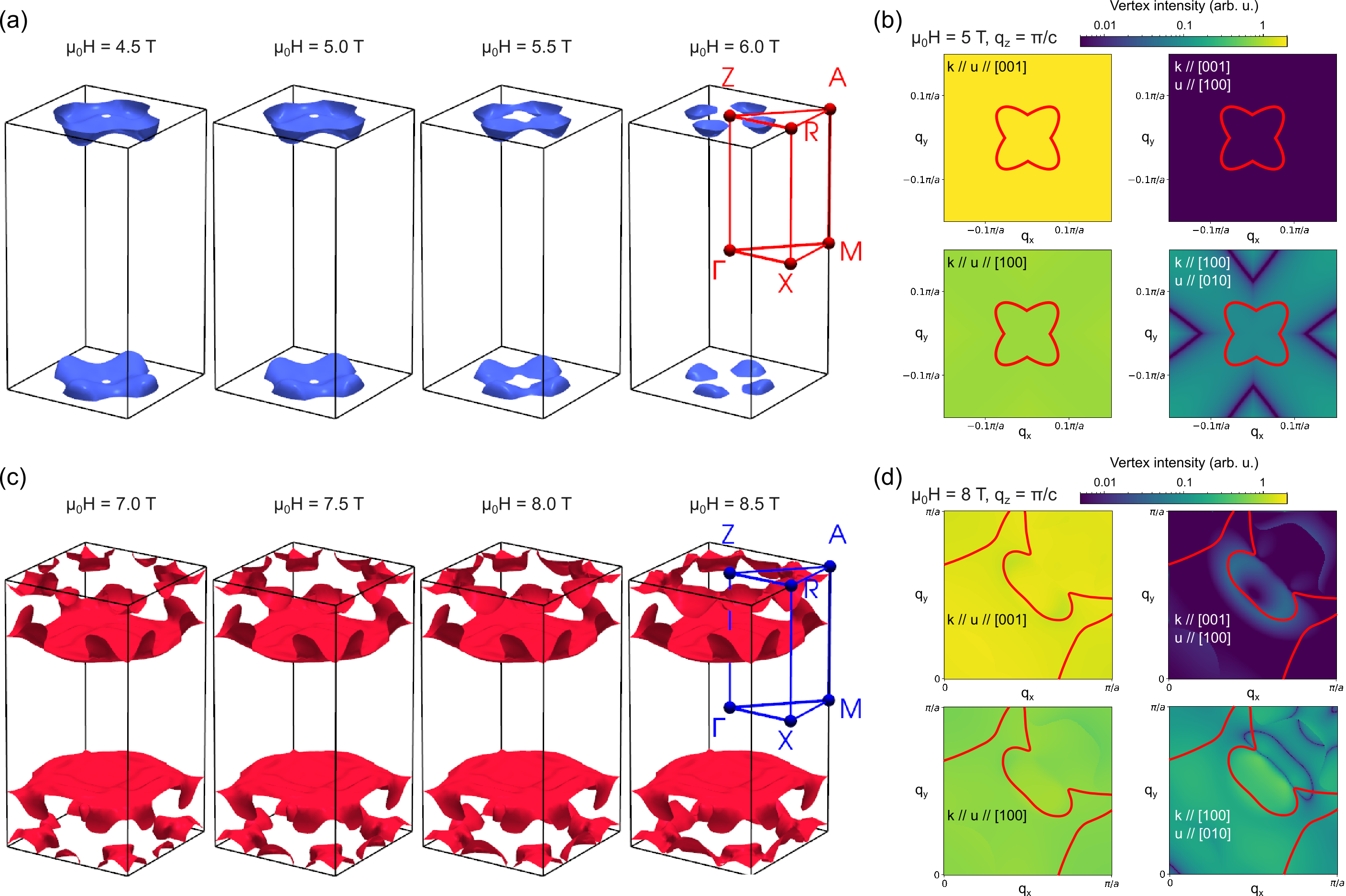}
    \caption{(a) Fermi surfaces for the band 8 at the vicinity of ETT No. 4. This ETT is characterized by the disparition of a small orbit around the $Z$ point of the Brillouin zone, associated to the magnetoacoustic quantum oscillations of the experimental frequency 34 T. (b) Ultrasound vertices $|M_{q}^{\beta = 8, \beta' = 8}|^2$ in the $q_z = \pi/c$ plane for the different acoustic modes, at $\mu_0 H = 5$ T. The orbit associated with the calculated quantum-oscillation frequency of 38 T is represented in red. (c) Fermi surfaces for the band 8 at the vicinity of ETT No. 5. This ETT is characterize by the fusion of two pockets in the $R \rightarrow A$ part of the Brillouin zone. (d) Ultrasound vertices $|M_{q}^{\beta = 5, \beta' = 5}|^2$ in the $q_z = \pi/c$ plane for the different acoustic modes, at $\mu_0 H = 8$ T. The three red lines represent the Fermi surfaces merging at ETT No. 5, in the upper and right edges of the BZ.}\label{fig7} 
\end{figure*}

Our ultrasound results revealed the presence of a small Fermi-surface orbit at low magnetic fields, corresponding to a quantum-oscillation frequency $F = 34$ T. This orbit vanish abruptly at the electronic-topological transition (ETT) No. 4, which occurs at a magnetic field of approximately $\mu_0 H \approx 6.2$ T. Quantum oscillations associated with this orbit are observed only for two acoustic modes: the longitudinal mode with $\mathbf{k}\parallel\mathbf{u}\parallel[001]$ and the transverse mode with $\mathbf{k} \parallel [100]$, $ \mathbf{u} \parallel [010]$. Using the Onsager relation $F = \hbar A^*/2\pi e$, where $\hbar$ is the reduced Planck constant and $e$ the electron charge, we extract an external cross-sectional area of $A^* = 0.32$ nm$^{-2}$ in the (001) plane. This corresponds to a filling fraction of 0.4 \% of the first Brillouin zone, whose total area is $A = 4\pi^2/a^2$ with $a=7.056 \times 10^{-10}$ m. This experimental value is in good agreement with our band-structure calculations, which predict a minimal orbit on band 8 at $q_z = \pi/c$ for a field of 5 T [shown in Fig. \hyperref[fig7]{\ref*{fig7}(a)}]. The predicted orbit occupies $0.45$ \% of the Brillouin zone, yielding a predicted frequency of 38 T.

Furthermore, our calculations predict the disappearance of this orbit at a magnetic field of 6 T [Fig. \hyperref[fig7]{\ref*{fig7}(a)}], consistent with the experimental observation. Above 5 T, the orbit size increases drastically before opening into 4 little pockets at 6 T, leading to ETT No. 4. Above 6 T, these 4 pockets disappear completely, and are not more visible at 10 T [Fig. \hyperref[fig1b]{\ref*{fig1b}(b)}].

\

We next discuss the symmetry selectivity of the observed magnetoacoustic quantum oscillations, which originates from the coupling between the cyclotron motion of electrons around the 38 T orbit and acoustic phonons. Figure \hyperref[fig7]{\ref*{fig7}(b)} shows the calculated mode-dependent electron-phonon coupling strength, represented by the ultrasound vertex intensity $|M_{q}^{\beta = 8, \beta' = 8}|^2$. The strongest coupling is predicted for the longitudinal mode with $\mathbf{k}\parallel\mathbf{u}\parallel[001]$, in agreement with our experimental observations. In contrast, the transverse mode with $\mathbf{k}\parallel[001]$, $\mathbf{u}\parallel[100]$ exhibits very weak coupling, again consistent with our observations. For the longitudinal mode with $\mathbf{k}\parallel\mathbf{u}\parallel[100]$, the calculations indicate a significant coupling between the phonons and the 38 T orbit. However, the absence of quantum oscillations for this mode suggests that a nonzero electron-phonon coupling is a necessary but not sufficient condition for the appearance of magnetoacoustic quantum oscillations. A reason to not observe the quantum oscillations could be a competing electron-phonon coupling in other parts of the Fermi surface, for example. Finally, for the transverse $\mathbf{k}\parallel[100]$, $\mathbf{u}\parallel[010]$, we predict a much stronger electron-phonon coupling compared to the other transverse mode, consistent with the experimental detection of oscillations in this configuration.

\subsection{ETT No. 5 at $\mu_0 H \approx 7.7$ T}

We now apply the same methodology to the ETT No. 5, detected in all acoustic modes at a magnetic field of approximately 7.7 T. Our electronic-structure calculations lead us to associate this transition with the formation of a large circular Fermi-surface pocket derived from bands 5 and 6 in the $q_z = \pi/c$ plane [Fig.  \hyperref[fig1b]{\ref*{fig1b}(b)}]. The detailed evolution of this pocket for band 5 is shown in Fig.  \hyperref[fig7]{\ref*{fig7}(c)}. We predict the formation of the large orbit at $\mu_0 H \approx 8.5$ T, in good agreement with the experimental value. At this field, two smaller pockets along the $R \rightarrow A$ path merge, leading to the ETT No. 5 [Fig. \hyperref[fig7]{\ref*{fig7}(c)}].

The corresponding ultrasound vertices are displayed in Fig. \hyperref[fig7]{\ref*{fig7}(d)}. The general hierarchy of coupling strength is similar to that observed for ETT No. 4: the longitudinal $\mathbf{k}\parallel\mathbf{u}\parallel[001]$ mode exhibits the largest vertex intensity, followed by the second longitudinal mode $\mathbf{k}\parallel\mathbf{u}\parallel[100]$, while the transverse modes show weaker coupling. This hierarchy can be understood from the factor $(\boldsymbol{\delta}\cdot \mathbf{u}_{k})(\boldsymbol{\delta}\cdot \mathbf{k})$ appearing in the $k\rightarrow 0$ limit of Eq. (\ref{eq10}), which usually is maximal for longitudinal modes $\mathbf{k} \parallel \mathbf{u}_{k}$. Consequently, electron coupling to transverse acoustic waves has often been neglected in heavy-fermion systems \cite{fulde1988theory,keller1990electron}.

However, our results for the transverse mode with $\mathbf{k}\parallel[001]$, $\mathbf{u}\parallel[100]$ clearly show that this mode is particularly interesting for the study of ETT in YbNi$_4$P$_2$. While its vertex intensity is negligible for ETT No. 4 [Fig. \hyperref[fig7]{\ref*{fig7}(b)}], a pronounced enhancement is found at ETT No. 5 [Fig. \hyperref[fig7]{\ref*{fig7}(d)}], in accordance with our ultrasound results. Finally, the second transverse mode $\mathbf{k}\parallel[101]$, $\mathbf{u}\parallel[010]$ also exhibits a relatively strong vertex at ETT No. 5, which also agrees with our experimental data.


\section{Conclusion}

We have conducted comprehensive ultrasound experiments on single-crystalline YbNi$_4$P$_2$ at low temperatures for magnetic fields applied along [001]. The sound-velocity data reveal a series of field-dependent anomalies, in agreement with previously reported thermodynamic and transport signatures of field-induced electronic-topological transitions (ETT). In addition, we have observed several quantum-oscillation frequencies in the sound velocity, providing direct evidence of a complex Fermi-surface topology. In particular, the abrupt disappearance of the 34 T oscillation at $\mu_0 H \approx 6.2$ T indicates the vanishing of a small orbit at ETT No. 4. Notably, the visibility of distinct ETT strongly depends on the acoustic mode, underscoring the key role of the symmetry properties of the electron-phonon interaction in determining the magnetoacoustic response.

To interpret these observations, we have developed a realistic electronic-structure model for YbNi$_4$P$_2$ that explicitly accounts for both the strongly correlated Yb $4f$ electrons via the Kondo effect, and the effect of external magnetic field. The phenomenological parameter of the theory was chosen to reproduce the experimentally measured Sommerfeld coefficient at low temperature. The resulting band structure displays strongly anisotropic Fermi surfaces and several field-induced electronic-topological transitions, in agreement with the experimental data. Furthermore, our calculations allow a direct correspondence between the experimentally detected anomalies and specific Fermi-surface reconstructions.

Finally, we introduced a microscopic model that incorporates the coupling between electrons and both longitudinal and transverse acoustic waves. This approach enables a quantitative determination of the ultrasound vertex in reciprocal space for each acoustic mode. We find that transverse modes, often neglected in heavy-fermion systems, can exhibit significant and selective coupling to specific regions of the Fermi surface near electronic topological transitions (ETT). This extended framework, thus, establishes ultrasound experiments as a sensitive probe of electron-phonon interactions and Fermi-surface reconstructions, offering enhanced resolution of ETT under extreme conditions of fields and temperatures.

\section{Aknowledgments}
We thank Zita Huesges and Toni Helm for fruitfull discussions. We acknowledge support from the Deutsche Forschungsgemeinschaft (DFG)
through SFB 1143 (Project No.\ 247310070), SFB/TRR 288 (422213477, project A03), 
W\"{u}rzburg-Dresden Cluster of Excellence on Complexity and 
Topology in Quantum Matter--$ctd.qmat$ (EXC 2147, Project No.\ 390858490),
as well as the support of the HLD at HZDR, member of the European
Magnetic Field Laboratory (EMFL).

\bibliography{biblio_JS_YbNi4P2.bib}

\end{document}